\documentclass[preprint,final]{aastex}
\usepackage{natbib}
\usepackage{float}
\shortauthors{Nataf et al.}
\usepackage{sidecap}
\shorttitle{Bulge Variables}
\shortauthors{Nataf et al.}
\usepackage{subfigure}
\usepackage{array,longtable}
\usepackage{color}
\definecolor{Blue}{rgb}{0.3,0.3,0.9}
\begin{document}

\title{HAT Discovery of 76 Bright Periodic Variables Toward the Galactic Bulge}
\author
    {
      D. M. Nataf\altaffilmark{1},
      K.~Z. Stanek\altaffilmark{1},
      G.\'A.~Bakos\altaffilmark{2},
    }
    \altaffiltext{1}{Department of Astronomy, Ohio State University, 140 W. 18th Ave., Columbus, OH 43210}
    \altaffiltext{2}{NSF Fellow, Harvard-Smithsonian Center for Astrophysics, 60 Garden St., Cambridge, MA~02138}
\email{nataf@astronomy.ohio-state.edu, kstanek@astronomy.ohio-state.edu, gbakos@cfa.harvard.edu}

\begin{abstract}
 We report on photometric I-band observations of 147 bright $(8<I<13)$ periodic variables toward the Galactic bulge including 76
 new discoveries. We used one of the HATnet telescopes to obtain 151 exposures spanning 88 nights in 2005 of an $8.4^\circ \times8.4^\circ$ field of view (FOV) approximately centered on $(l,b) = (1.73, -4.68)$.
We observed the galactic bulge in 2005 as part of a microlensing feasibility study
 \citep{2009AcA....59..255N}, and here we discuss the periodic variables we found in our data. Among our discoveries we count 52 new eclipsing binaries and 24 other periodic variables. 
\end{abstract} 


\section{Introduction}

The Galactic bulge is one of the most studied  regions of the sky, with recent surveys focused on a broad array of different observables such as deep optical sources \citep{2003AcA....53..291U}, the near IR in the context of globular clusters \citep{2007AJ....133.1287V}, hard X-rays \citep{2007A&A...466..595K}, $\gamma -$rays \citep{2008Natur.451..159W}, and extinction toward the Galactic center from the optical through the mid-IR \citep{2008ApJ...680.1174N, 2009ApJ...696.1407N}, among others. One of the corollary benefits of our small-aperture microlensing feasibility study  \citep{2009AcA....59..255N} was the power of a time-series of exposures in a relatively unexplored part of the observational parameter space. What we have with our time-series is a snapshot of bright sources, with apparent magnitudes ranging from  8 to 13 in $I$. As the distance modulus to the bulge is approximately 14.5 \citep{1998ApJ...503L.131S,2006ApJ...647.1093N, 2006A&A...450..105B, 2008A&A...481..441G, 2009A&A...498...95V, 2009MNRAS.398..263M, 2009MNRAS.399.1709M} we will be looking at primarily bright stars: bulge giants and brighter main-sequence stars on sightlines toward the bulge but at intermediary distances. This selection effect for brighter stars is amplified by the high extinction toward the bulge. \citep{2004MNRAS.349..193S}, in his analysis of the OGLE-II bulge data, found an extinction of $A_{I}$=$(2\pm1.5)$. Our observational window overlaps with parts of OGLE-II but also likely include sightlines of much greater extinction. 

\newpage
\begin{figure*}[p]
\begin{center}
\includegraphics[totalheight=0.81\textheight]{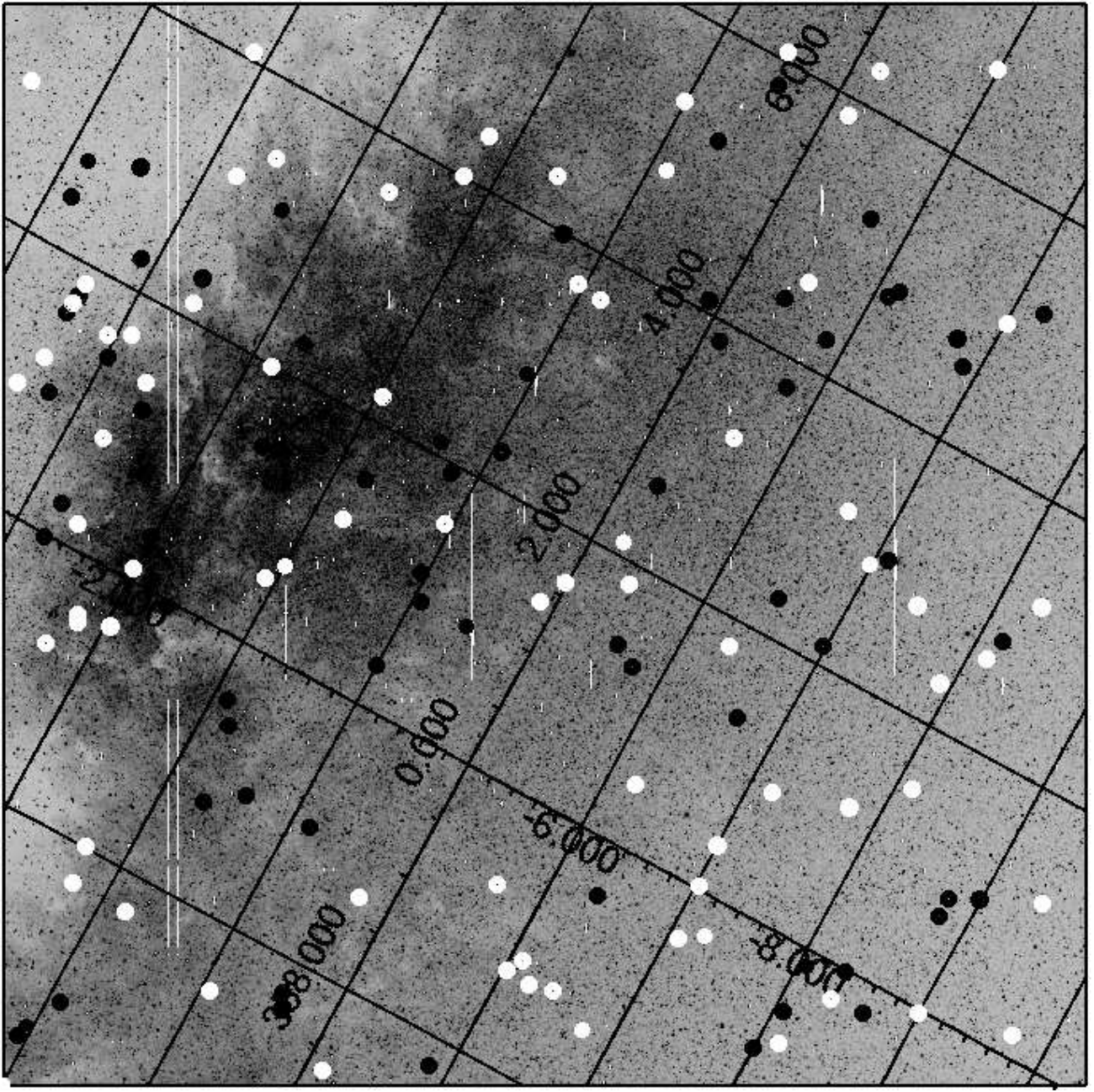}
\end{center}
\caption{Image of our field, with galactic coordinates shown. Of our 147 detected periodic variables, the 71 previously known are shown in black, the 76 new periodic variables are shown in white}
\label{ReferenceImage}
\end{figure*} 
\clearpage

Both these populations, the foreground stars toward the bulge and the brightest bulge giants, are independently interesting. With recent findings that the extinction law is a strongly-varying function of galactic position \citep{2003ApJ...590..284U,2004MNRAS.349..193S,2005ApJ...619..931I,2009ApJ...707...89G,2009ApJ...707..510Z}, better understanding of extinction toward the bulge may require anchoring extinction at various points between us and the bulge. Meanwhile, a curious discrepancy has manifested itself between the metallicities of bulge giants and microlensed dwarf stars \citep{2010ApJ...709..447E,2010ApJ...711L..48C} - there may be constraints to be gleaned by having a more complete inventory of the brightest bulge giants.

 Our data was taken during the 2005 bulge season, when we obtained 151 exposures. This was done to investigate the idea of \citet{1998ApJ...497...62G}, that it is feasible for a single wide-field instrument to take high-cadence observations of the brightest stars toward the bulge. That would represent a potential boon for gravitational microlensing, as the highest-magnification microlensing events have an impressive scientific track record of informing exoplanetology, (\textit{e.g.} \citet{2010ApJ...720.1073G}). We reproduced several 2005 OGLE-III events toward our field \citep{1994AcA....44..227U}, thereby demonstrating the feasibility of a Small Aperture Microlensing Survey (SAMS) and proving the concept of \citet{1998ApJ...497...62G}, the reader is referred to \citet{2009AcA....59..255N} for a more complete analysis.  In this paper, we investigate an expected side-benefits of a SAMS, the prospect of acquiring a complete census of the brightest periodic variables toward the bulge. Following extensive reductions, analysis and inspection of our 115,624 point sources, we arrived at a sample of 147 periodic variables. We compared these to the GCVS \citep{1997BaltA...6..296S}, ASAS \citep{2004AN....325..553P} catalogs, and OGLE-II variable star catalog \citep{2002AcA....52..129W} and found that 76 (52\%) were previously undiscovered. At least 52 of these are eclipsing binaries. That we obtain such a high yield in the most heavily studied region of the sky with a low cadence and observational span tells us that a large amount of periodic variables remain to be discovered. The yield would have been even higher were it not for overlap with the ASAS variable star search, its southern hemisphere searches between 12th and 18th hour \citep{2004AcA....54..153P}
and between 18th and 24th hour \citep{2005AcA....55...97P}. 

We discuss our data and reduction in section \ref{section:Data}, our method of periodic variable selection in section \ref{section:OVS}, we present our results, discuss some interesting cases and our conclusions in \ref{section:DiscussionNConclusion}. The phased light curves of our new periodic variables along with tables of all observed variables are appended at the end of the paper.\label{section:Introduction}
\section{Data and Reduction}
\label{section:Data}

We closely follow the approach of \citet{2004AJ....128.1761H}. We used the ISIS image subtraction package \citep{1998ApJ...503..325A,2000A&AS..144..363A}  to reduce our data, and the DAOphot/ALLSTAR package \citep{1992JRASC..86...71S} to construct our source list. Below we list only key points and a generalized overview. A more detailed summary of our data and reduction can be found in \citet{2009AcA....59..255N}.

\begin{figure}[ht]
  \begin{center}
\includegraphics[totalheight=0.4\textheight]{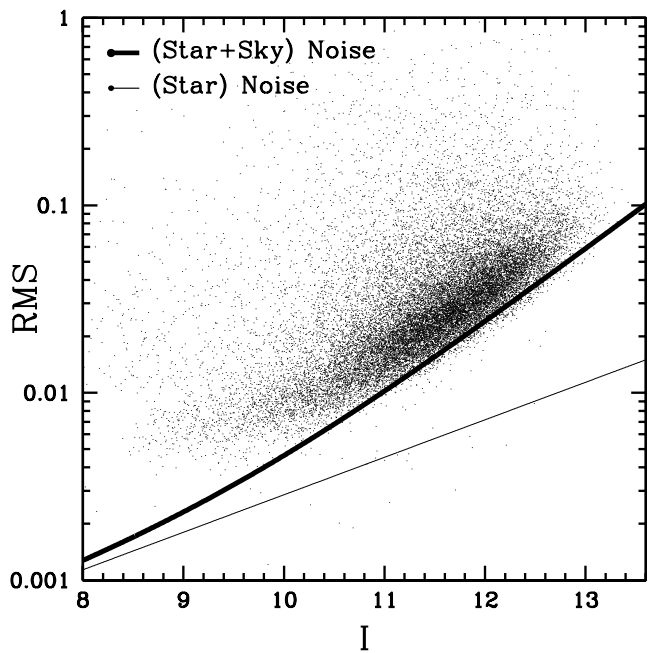}
\caption{The RMS for our light curves is shown as a function of mean magnitude. The theoretical photon-noise limit plotted in the thin black line, and the estimated combined noise from source and sky is plotted in the thick black line.}
  \end{center}
  \label{fig:RMSvsMean}
\end{figure}

Our observations were taken over an 88 day span during the 2005 Bulge season, yielding 151 exposures over 61 distinct nights to image a $8.4^{\circ}\times 8.4^{\circ}$ FOV. We used one of the HAT telescopes, HAT-9, located on Mauna Kea, Hawaii. A 11~cm diameter Cannon f/1.8L lens was used to image onto an
Apogee AP10 2K$\times$2K CCD. The resulting pixel scale is
$14\arcsec$. A detailed description of the equipment, observing
program and preliminary CCD reductions can be found in \citet{2004PASP..116..266B}. Our photometric precision is shown in Figure \ref{fig:RMSvsMean}. Magnitude calibration was done by comparing instrumental magnitudes to photometry of OGLE \citep*{1997AcA....47..319U} objects in our field with I$\le$12.5. As this is one of the densest fields in the sky blending is a significant concern for the faintest sources, and those with $I>12$ are more likely to have their brightness overestimated. 

We performed astrometry on the selected reference frame as described in Pal\& Bakos (2006).  We sought the astrometric solution in the form of fourth order polynomials connecting the X and Y pixel coordinates with the
Cartesian $\xi$ and $\eta$ (arc-projected) coordinates around the field
center, as taken from the 2MASS catalog  \citep{2006AJ....131.1163S}.  The pixel coordinates of the stars were derived by DAOPHOT.  For the brightest sources these pixel coordinates were accurate to $\sim 0.03$\,pixels.  The astrometric solution was found by triangle matching, and then refined by matching the $\sim
10,000$ brightest stars on the frame with the 2MASS catalog.  The r.m.s.~around the best fit is 0.22 pixels (3.2 arcseconds).  We then used this astrometric solution to transform all of our (X, Y) coordinates to the World Coordinate System (that of 2MASS, epoch 2000.0, equinox 2000.0), and matched the sources against 2MASS.  For stars with a clear match from 2MASS we kept the 2MASS coordinates.  For sources with ambiguous or no matches, we keep our projected (X, Y coordinate based) RA and DEC coordinates. 

\begin{figure}[H]
\begin{center}
\includegraphics[totalheight=0.59\textheight]{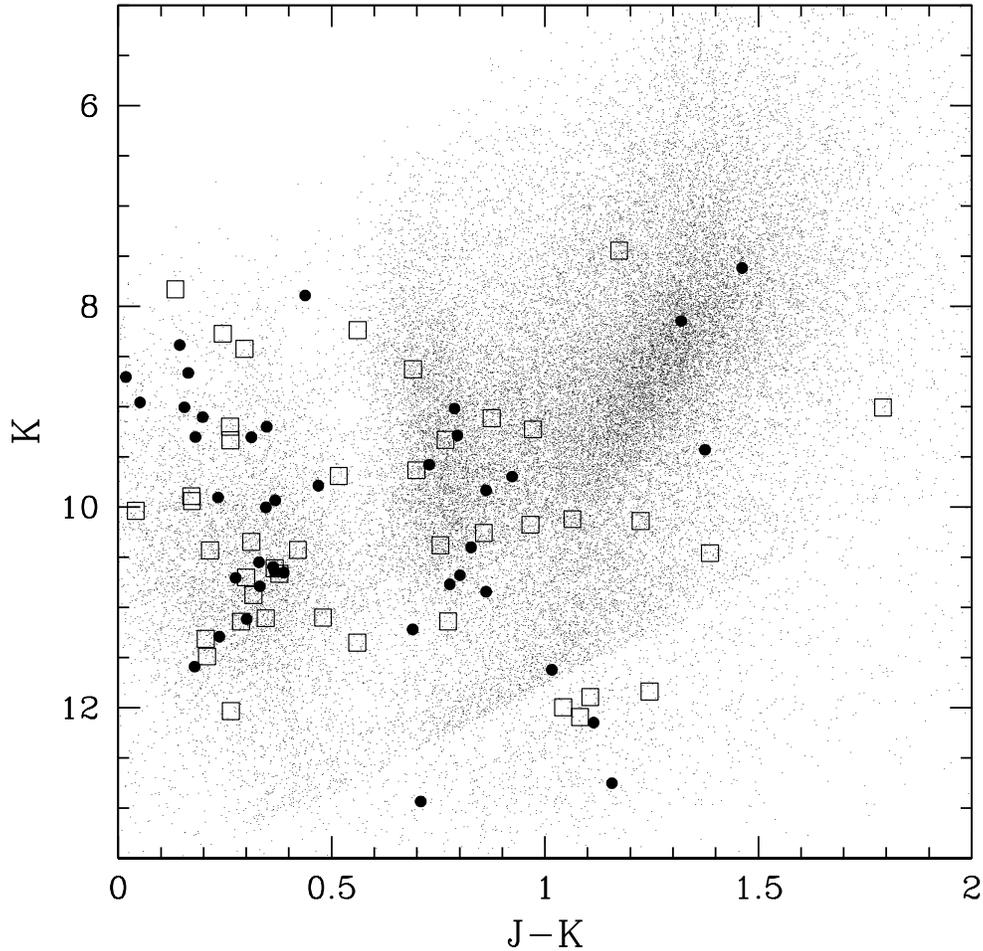}
\caption{2MASS Color-Magnitude diagram in $(K,J-K)$ for our $\sim$47,000 matches. The 42 known periodic variables with 2MASS matches are shown as the empty squares, and the 42 new periodic variables with 2MASS matches are shown as the filled circles. }
\end{center}
\label{2MASSCMD}
\end{figure}

The $(K,J-K)$ color-magnitude diagram for our matches is shown in Figure \ref{2MASSCMD}. A few trends are apparent in that figure. First, most of the periodic variables detected, shown as the empty squares (previously known) and the filled circles (new discoveries) are among the bluest stars in CMD, indicating that they are foreground disk stars rather than background bulge stars. Second, the entire CMD spans a broad range in color, $0.0 \lesssim (J-K) \lesssim 2.0$, almost continuously, indicating the high degree of differential reddening. We note that the overdensities in the CMD are not due to red clump stars, but are an artifact of the rapidly declining detection sensitivity to stars fainter than $I=12$.

\section{Periodic Variable Selection}
\label{section:OVS}

We ran  the Vartools implementation \citep{2008ApJ...675.1468H} of an analysis of variance (AoV) period search developed by \citet{1989MNRAS.241..153S} and \citet{2005ApJ...628..411D} on our 115,624 sources. We searched for periods between 0.1 and 100 days. 

\begin{figure}[ht]
  \begin{center}
    \subfigure[All     lightcurves]{\label{fig:edge-a}\includegraphics[scale=0.4]{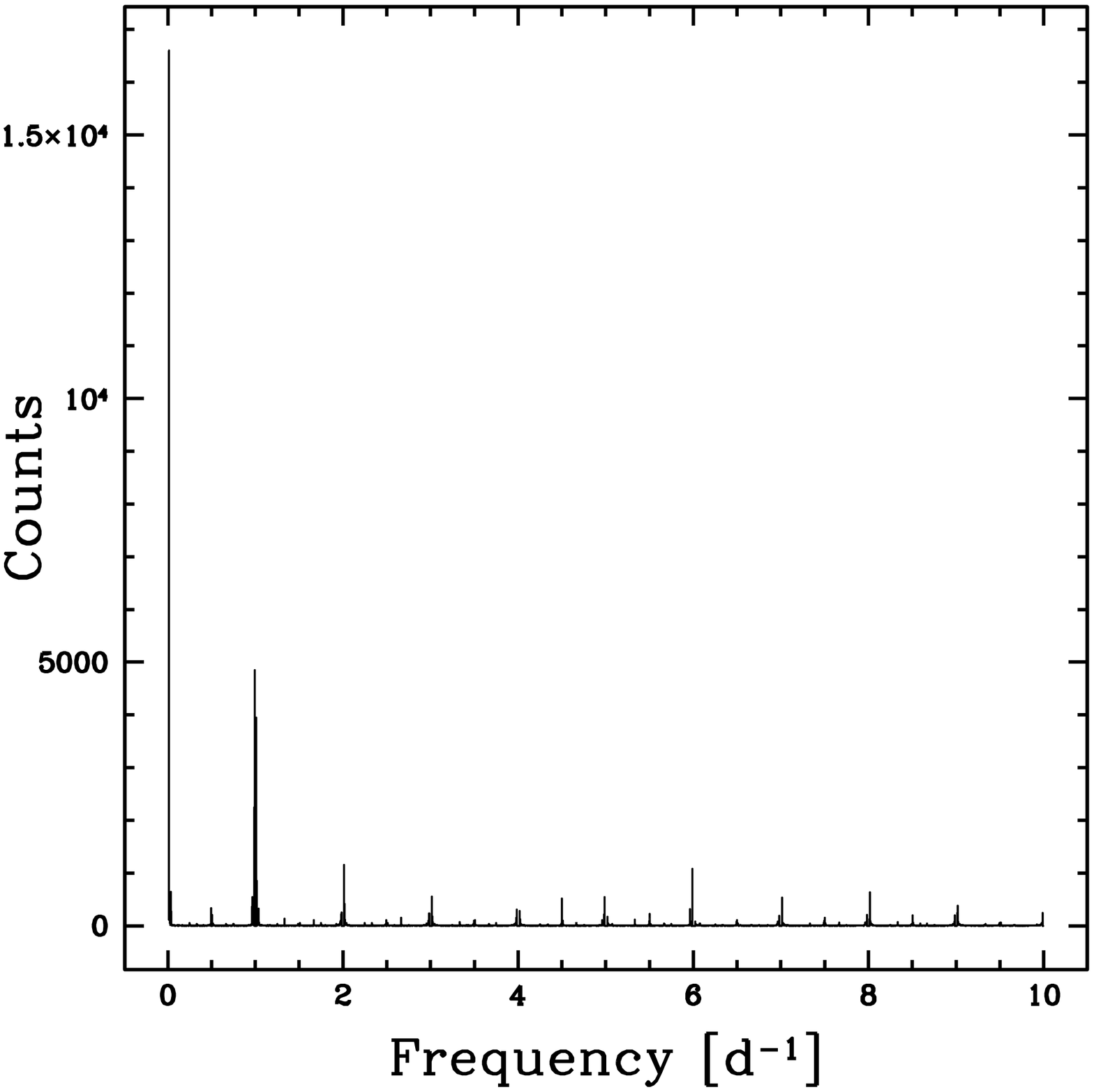}}
    \subfigure[After bins with n$>$25 were  removed.]{\label{fig:edge-b}\includegraphics[scale=0.4]{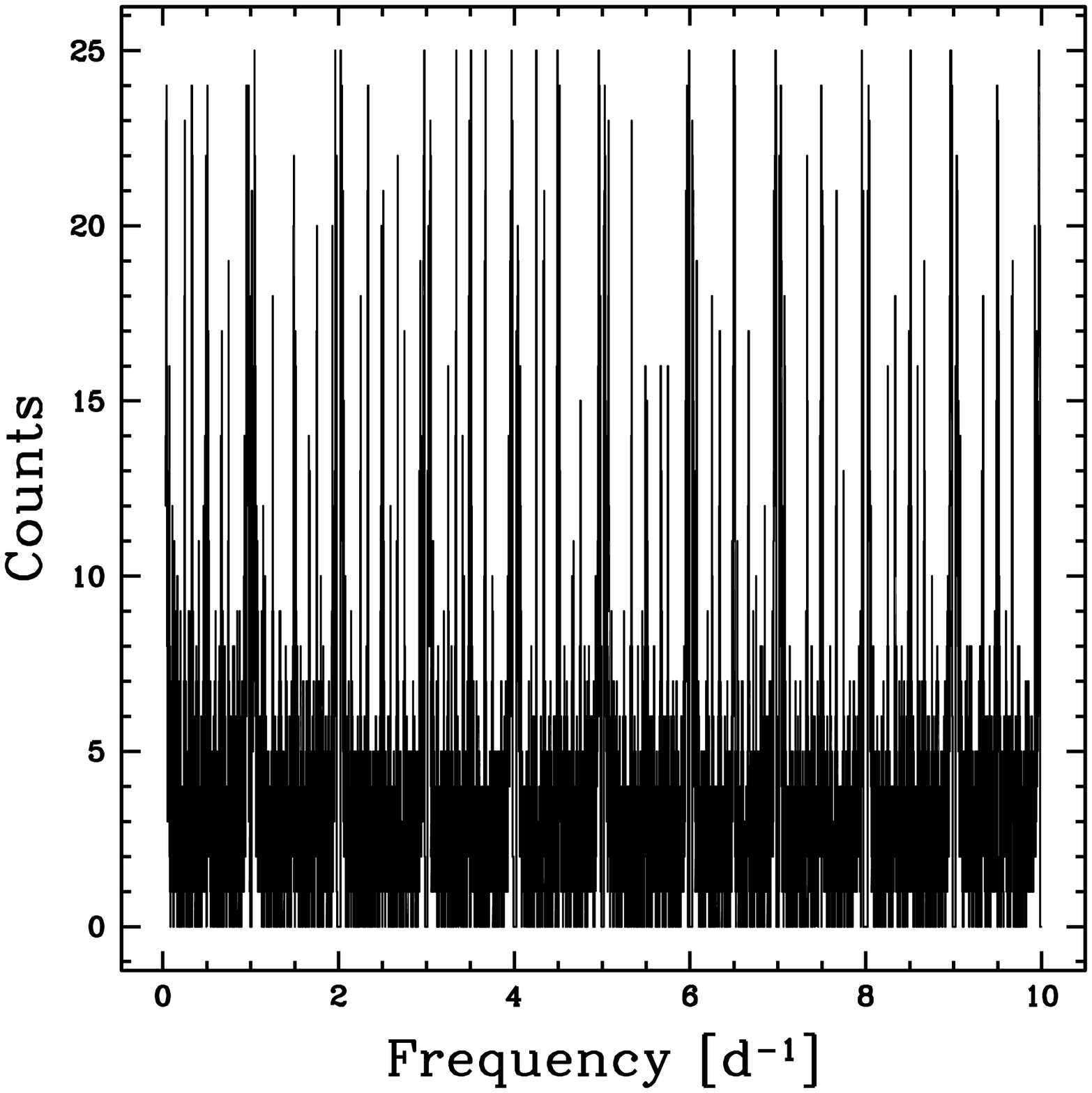}} 
  \end{center}
  \caption{Distribution of best-fit frequencies as selected by AoV}
  \label{fig:edge}
\end{figure}

At our sampling - 151 observations spread over 88 nights - aliasing is a significant issue. Many of our lightcurves will exhibit strong Fourier signals for specific periods, particularly at harmonics of 1 day. At the same time, the underlying population of truly periodic variables should have their periods smoothly spread over a periodogram, which should still be the majority of frequency space. Exploiting this, we binned our best-fit periods into 9990 frequency bins covering a range between 0.1 and 100 days. We kept all lightcurves with an empirically-selected variability alarm AoV$>$ 3.33 that had their best-fit period in bins with 25 or fewer lightcurves. That is a generous cutoff, as a uniform distribution would lead to an expectation of only 10 lightcurves per bin. Figure \ref{fig:edge} demonstrates the principle.

We discarded 491 of our 9990 bins, 4.9\% of frequency space, which in turn contained 80,626 of our 115,624 lightcurves. As such the best variable-extraction performance we could achieve would be $\sim$95\% completeness. However, for any realistic cadence, photometric precision and temporal baseline there will be variable classes at the threshhold of detection, and these have a higher probability of being shifted into ``noise'' bins, lowering our expected recovery rate. We manually inspected the 1,588 lightcurves that were in good bins and had an AoV$>$3.33, an very low cutoff but rendered possible with the now parsed population of lightcurves. We selected 254 sources, ran a finer AoV search on this smaller sample, and then kept 204 variables.

A systematic issue which crops up in dense fields is that of ``variability ghosts''. With blending, a single variable star, if sufficiently bright, may make variables out of multiple nearby stars that are separated by an angular distance smaller or comparable to the width of the PSF. We corrected for this by searching for variable sources within our final list that were nearby to one another, and when the mode of variability was the same we kept the source with the higher amplitude of variation in photon flux. 

Figure 4 is a histogram displaying how the remaining lightcurves populated the bins. With $\sim$35,000 lightcurves and $\sim$9,500 bins, one would expect an approximately Poisson histogram with a mean of 4, \textit{i.e.} the typical frequency bin would contain the best-fit frequencies to 4 lightcurves. As the remaining distribution is not quite Poissonian one can conclude that even the remaining sample is not ``fully cleaned'' of bad frequency bins. The Poisson expectation is that virtually no frequency bins contain more than 10 lightcurve matches, whereas we obtain a fair number. Likewise, we also obtain an excess of frequency bins with only 1, 2 or 3 lightcurve matches. We nevertheless chose the cutoff of 25 based on our observations that less generous cutoffs cause the frequency-bin rejection fraction to rise much higher than 5\%. 

We used the Lomb-Scargle algorithm \citep{1976Ap&SS..39..447L,1982ApJ...263..835S,1989ApJ...338..277P,1992nrca.book.....P} as a second means of periodic variable acquisition, in case it could find a few variables missed by AoV. We took the sources which had an AoV$>$3.33 but were in congested frequency bins. We first applied a $J_{S}>1.2$ cut \citep{1996PASP..108..851S}, a measure of correlated variability, thus keeping 11,731 sources. We then ran the Vartools implementation of LS on periods between 0.1 and 44 days, where 11,516 lightcurves (98.2\%) were removed for having their best-fit frequency in 49 of the 2000 (5.6\%) frequency bins. This left 140 lightcurves which we directly inspected visually. Even then our sample was contaminated. However, it was sufficiently small to be inspected manually. Only 2 of these 140 lightcurve candidates were added to our periodic list, indicating either a high rate of completeness from our AoV search or the fundamental similarity of the two approaches in this observational context.

\begin{figure}[ht]
  \begin{center}
\includegraphics[totalheight=0.4\textheight]{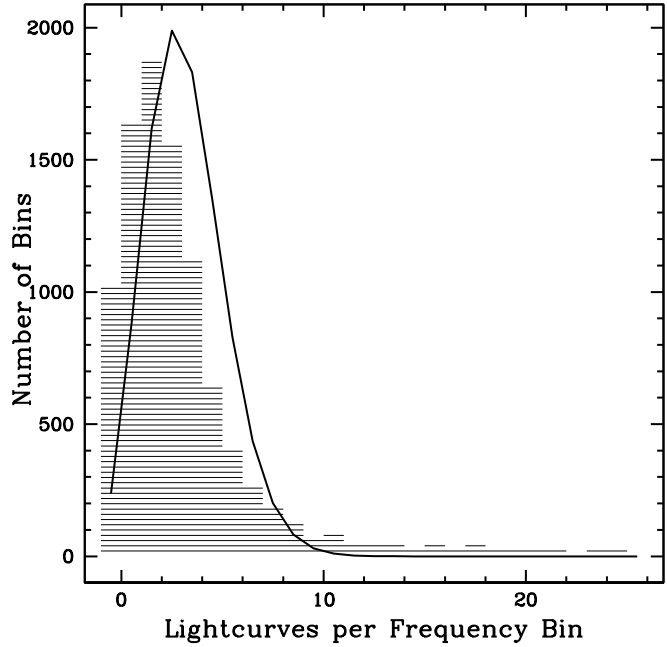}
\caption{Distribution of the lightcurve AoV periods in frequency space. A histogram of the frequencies per bin once bins with n$>$25 were removed. In red we plot the theoretical expectation based on Poisson statistics.}
  \end{center}
  \label{fig:distrib}
\end{figure}

\begin{figure}[H]
  \begin{center}
    \subfigure[]{\label{fig:edge3-a}\includegraphics[scale=0.4]{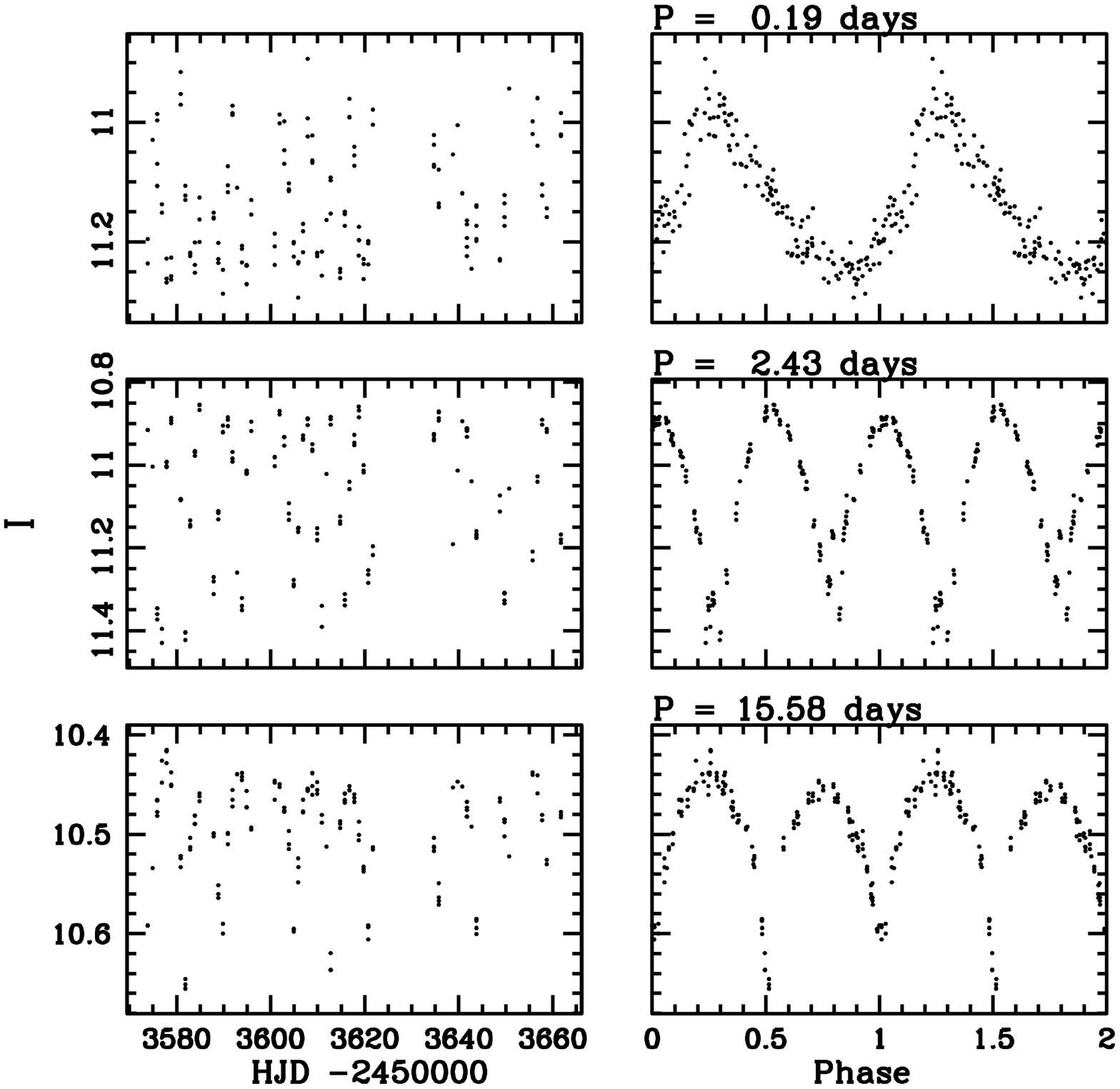}} 
    \subfigure[]{\label{fig:edge3-b}\includegraphics[scale=0.4]{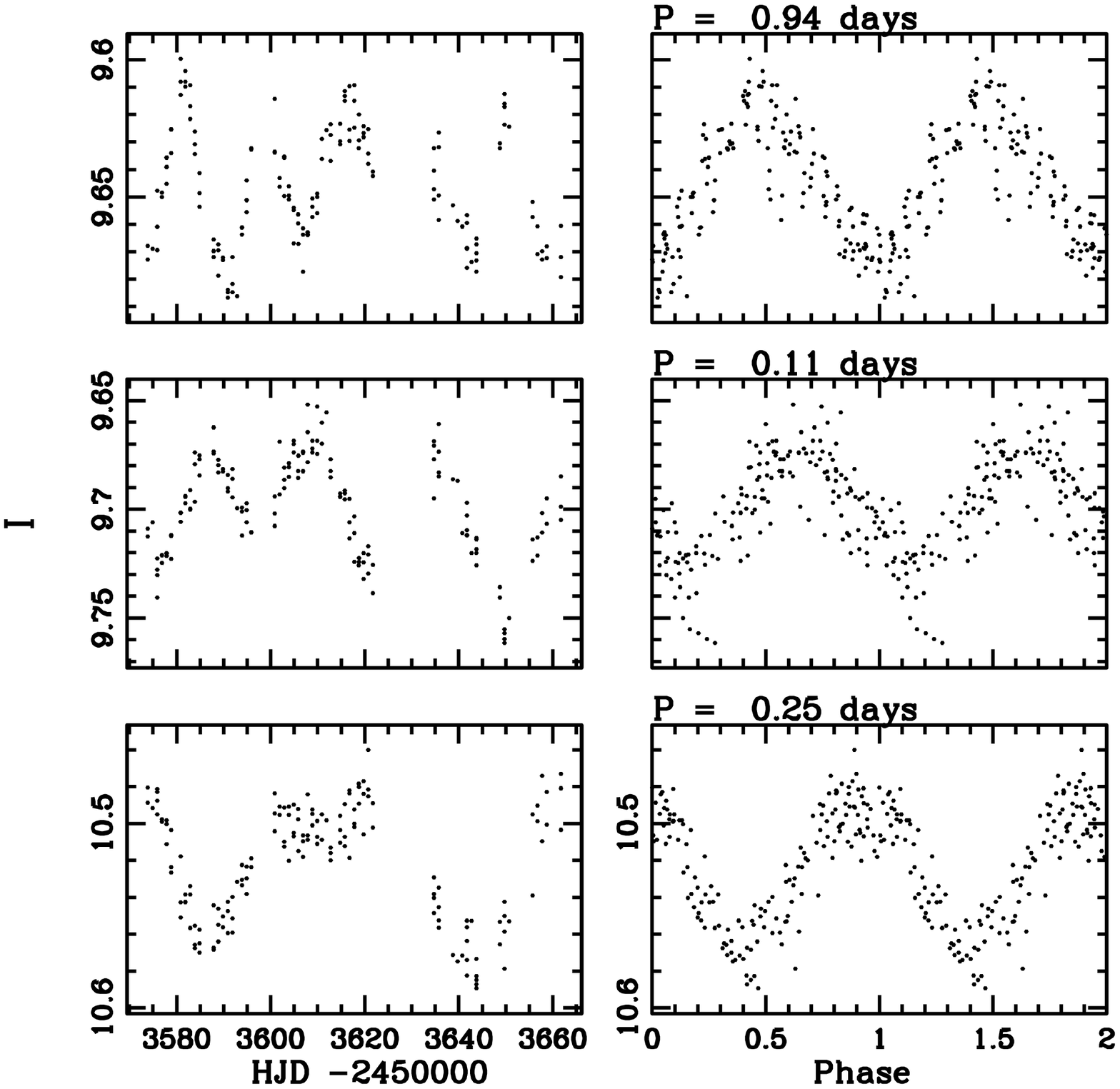}}
  \end{center}
  \caption{On the left, variable stars securely identified as periodic variables. To the right, 3 variable stars well-fit by a likely spurious period due to sampling. The unbinned lightcurves are shown on the left for both sample sets. }
  \label{fig:edge3}
\end{figure}

The population of bad bins is much more extensive than merely the harmonics of one day. A possible explanation for this is that observations are not really spaced exactly one day apart due to factors such as changing sunset/sunrise time, and the length of the exposure time itself. We had observations on 66 of the 88 nights, with a varying amount of observations per night and of nights between observations. Contamination is thus spread to many other bins, spread densely over the frequency space. 


Even after removing bad frequencies, many sources can still be well-fit by a periodic curve in spite of not being true periodic variables, they are likely false positives. We plot 3 cases of such variables in Figure \ref{fig:edge3-b}. It is possible that these sources do possess intrinsic periodicity, perhaps due to rotational brightness modulation, but we could not be sure from our lightcurves alone. Figure \ref{fig:edge3-b} are a representative sample of more solid finds. Even their unphased lightcurves imply periodicity.

One potential concern is that since the sampling in the time-series generates this issue, perhaps most or nearly all of our periodic variables are spurious? This risk was taken seriously and this is why we opted for a smaller but secure data set rather than a larger but likely contaminated data set. For eclipsing variables, nature has provided an additional means of security - their characteristic shape, which is much less likely to be stochastically mimicked than that of a pulsator. A risk remains some periods could be off by a factor of two,  when only one eclipse is visible. In those cases, we chose to assume the binaries were twins, and thus to divide the period into two, rather than that the other possibility, that of a very lopsided surface temperature ratio. Ultimately, a multi-band photometric time series is the best way to resolve this degeneracy.

For our other periodic variables, which are most likely pulsators, we note that we have approximately the same fraction of new discoveries as we do for the more secure eclipsing variables. Some pulsators, such as RR Lyrae, also have characteristic shapes. As seen in Figure \ref{fig:edge3}, lightcurves with more recognizable shapes were also fit with periods further removed from the 1-day frequency harmonics.

\section{Discussion \& Conclusion}
Our find of 76 new periodic variables implies there remains a large number of undiscovered periodic variables toward the galactic bulge, specifically in the brighter magnitude ranges which we explored. As we only had 151 data points, grouped closely together on 66 nights which themselves spanned 88 nights, we expect there to be many more variables which would be extractable with a longer baseline and/or a higher cadence.

\begin{figure}[p]
  \begin{center}
    \subfigure{\label{fig:LogPeriod-a}\includegraphics[scale=0.52]{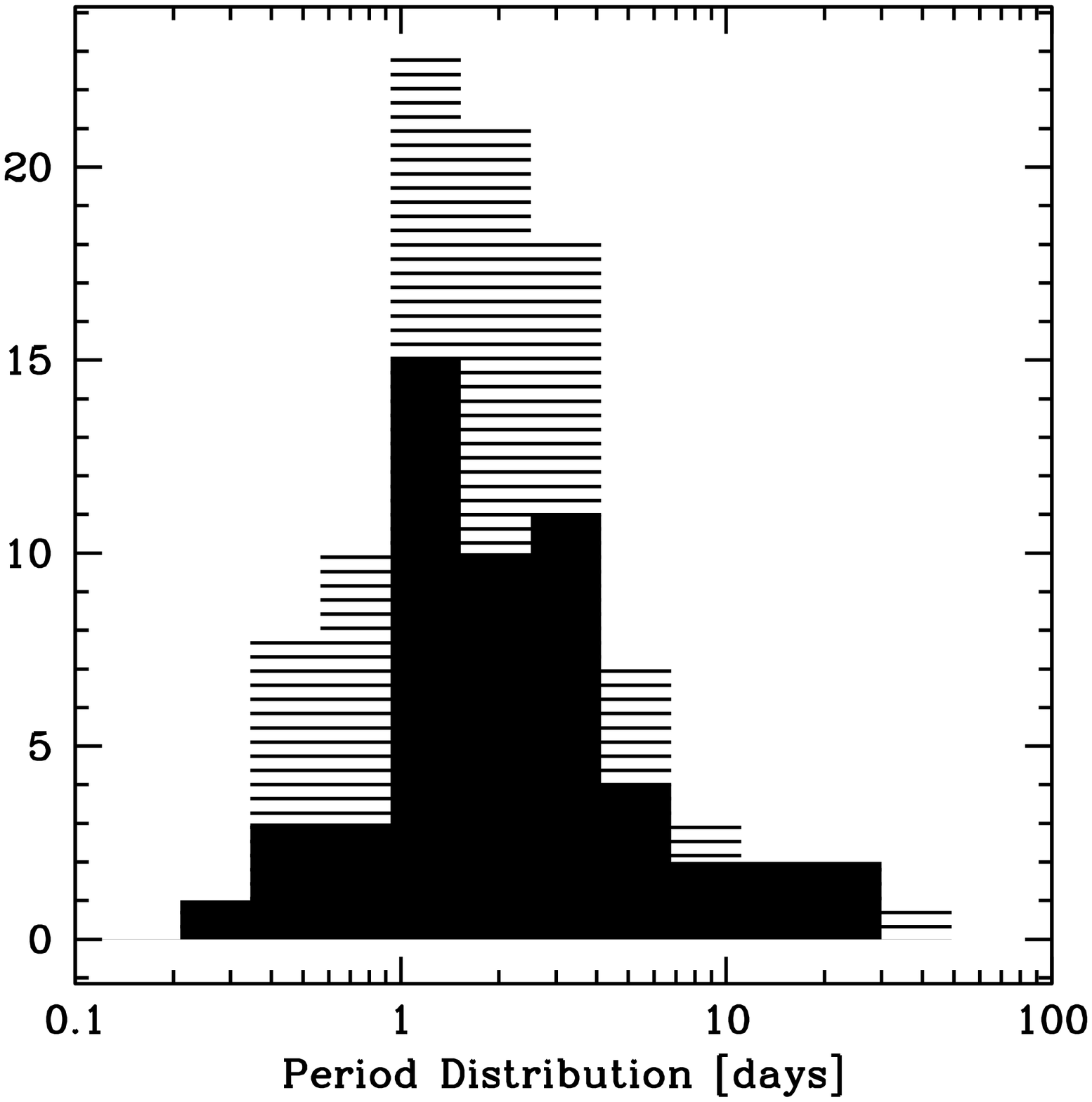}}
    \subfigure{\label{fig:LogPeriod-b}\includegraphics[scale=0.52]{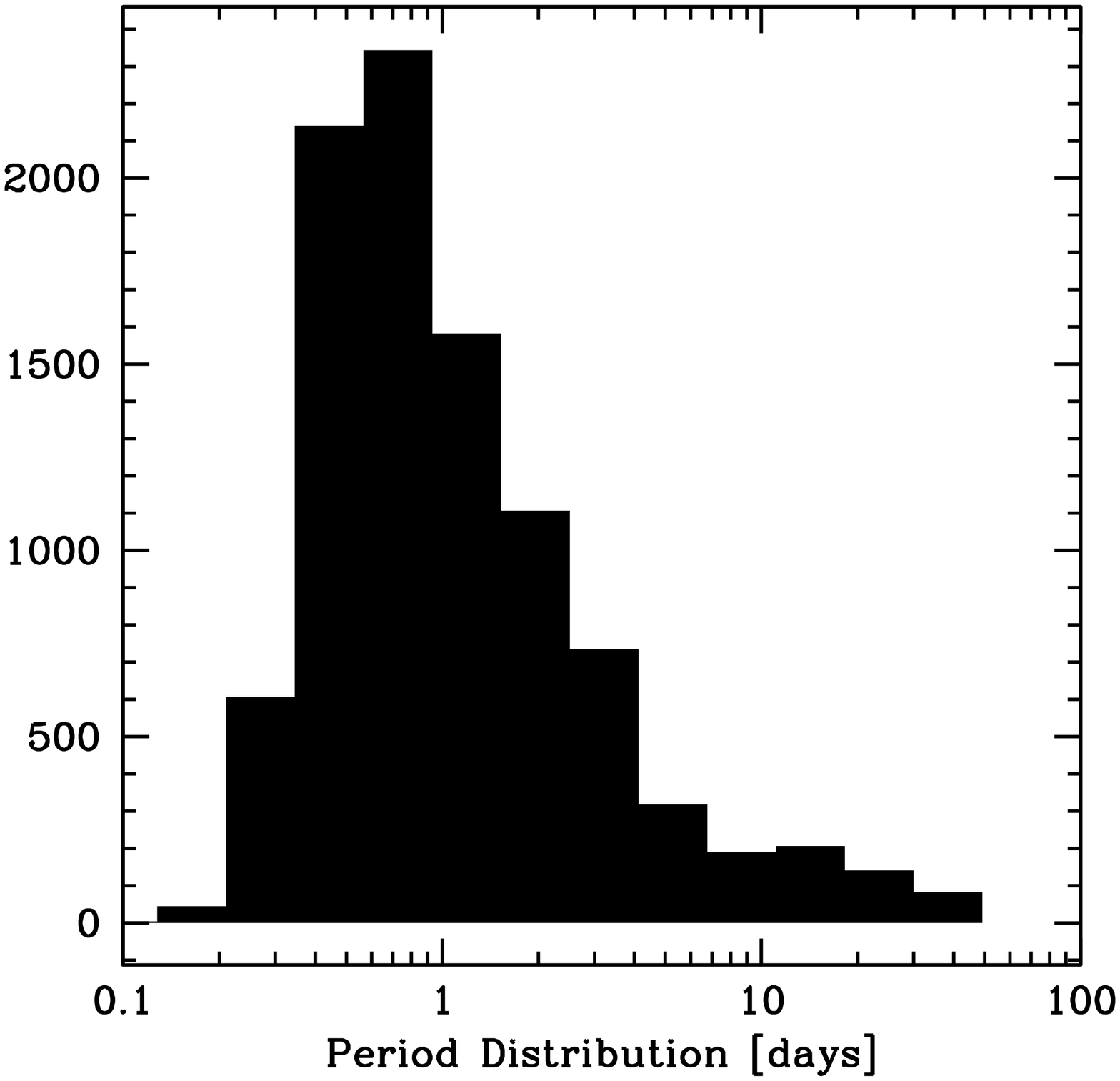}} 
  \end{center}
  \caption{Top: The distribution of periods of our final sample of our 52 previously unknown eclipsing variables is shown as the shaded histogram. Adding the eclipsing binaries previously known from ASAS and GCVS yields a sample of 96, seen in the striped histogram. Bottom: We show as comparison the distribution of 9,501 of the 10,862
OGLE-II eclipsing binaries \citep{2005ApJ...628..411D} that fall within our period range. }
  \label{LogPeriod}
\end{figure}

In Figure \ref{LogPeriod} we show the period distribution of our eclipsing binary population and we compare to that of 10,862 OGLE-II eclipsing binaries cataloged by \citet{2005ApJ...628..411D}\footnote{Obtained from: http://www.iop.org/EJ/article/0004-637X/628/1/411/61389.html} for comparison. Our findings are similar with their distribution, which is not obviously expected. The apparent magnitude range of the \citet{2005ApJ...628..411D} sample is approximately $12<I<20$, with a median magnitude of $\sim$17 - with our range of $8<I<13$ we are probing a brighter and thus different population. They are likely probing a different range of stages of stellar evolution, and thus possibly of different initial masses and metallicities, depending on the history of star formation in the bulge. We do note some differences. The mode of our distribution is at $\sim$1 day, whereas \citep{2005ApJ...628..411D} lies at $\sim$0.5 days. More observations are needed to ascertain if this if a true property of foreground disk binaries and/or bulge binary giants rather than a statistical fluke. A more complete sample of the bright eclipsing binaries could potentially impose tight constraints on binary evolution models and of the star formation history in the bulge. Using numerical models,  \citet{2007ApJ...669.1298F} investigated how a binary population period distribution would evolve after 10 billions years from effects such as tidal and 3-body Kozai interactions. They predicted a surge of binaries with periods between 1 and 10 days from 3-body interactions. The larger the fraction of binaries in hierarchical systems with a distant ternary companion, the more significant this effect is expected to be.

\begin{figure}[H]
\begin{center}
\includegraphics[totalheight=0.4\textheight]{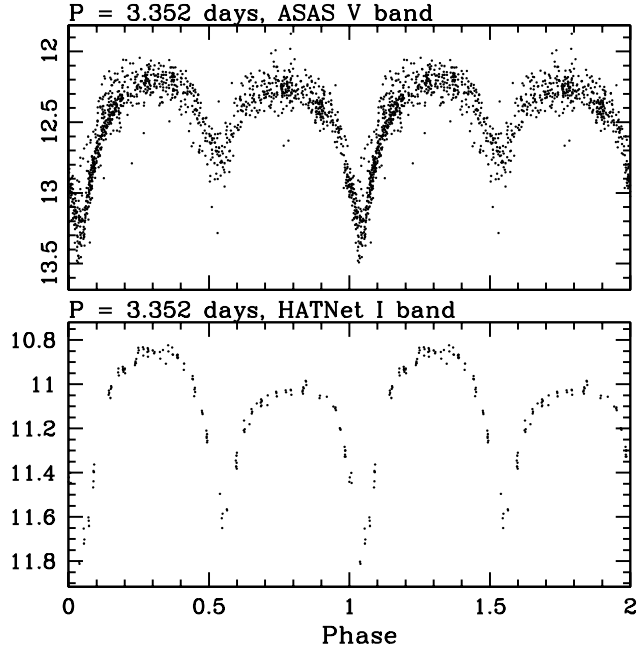}
\caption{An interesting eclipsing binary, HAT-622-37664, located at ($\alpha,\delta$)= (18:24:29,-32:25:57). On top, we display the ASAS data, observing a different shape to the light curve in the shorter wavelength V-band, where the amplitude of the O'Connell effect is reduced.}
\end{center}
\label{InterestingPeriod}
\end{figure}

We comment on an interesting variable. In Figure \ref{InterestingPeriod}, we show a known periodic variable, ASAS 182429-3225.6. The ASAS data plotted in green is from one of their 4 cameras, and is in V-band. Both the ASAS plot and our plot exhibit the O'Connell effect, a light asymmetry between successive maxima in an eclipsing binary\citep{1951PRCO....2...85O}. It appears more pronounced in I-band. Explaining the O'Connell effect has been one of the challenges in close binary star models, and efforts continue to do so, using models such as captured circumstellar material  \citep{2003ChJAA...3..142L} and starspot activity \citep{1960AJ.....65..358B,1989ApJ...343..909L,1990MNRAS.247..632B}. Some observational and statistical investigation of the effect continues, \citet{2007MNRAS.378..757P} searched for correlations between the amplitude of the O'Connell effect and period change rates in eclipsing binaries. The poorly understood variation of the amplitude of the O'Connell effect with bandpass may cloud such studies. As the ASAS survey \citep{2004AN....325..553P} provides time-series for large-swaths of the bulge in $V$, and OGLE \citep{2008AcA....58...69U} has its highest cadence in $I$, a SAMS instrument might maximize its complementary potential by taking data in $J$ or $H$.

Using wide-field I-band observations, a broad PSF and image subtraction we have demonstrated that there remains significant numbers of bright periodic variables of interest not detected in the direction toward the Galactic Bulge. We count 76 discoveries, of which at least 52 are eclipsing binaries, and the rest are likely pulsating variables. Our observational parameters are very similar to what we suggested in  \citep{2009AcA....59..255N} as an accompanying survey to current microlensing surveys, which demonstrates such a survey would have the corollary benefit of increasing the number of known periodic variables. As these occupy a specific observational parameter space - higher brightness - they may correlate to a specific region of physical parameter space - giant stars.\label{section:DiscussionNConclusion}

\acknowledgements{
DMN is partially supported by the NSF grant AST-0757888.

HATNet operations have been funded by NASA grants
NEG04GN74G, NNX08AF23G and SAO IR\&D grants. We thank J.D. Hartman for
assistance with use of the Vartools package, and G. Pojmanski for his
``lc'' program. We made use of the Simbad database, operated in Strasbourg, France by the \textit{Centre de Données astronomiques de Strasbourg}.}

This publication makes use of data products from the Two Micron All Sky Survey, which is a joint project of the University of Massachusetts and the Infrared Processing and Analysis Center/California Institute of Technology, funded by the National Aeronautics and Space Administration and the National Science Foundation.


\newpage
\begin{figure}[H]
\includegraphics[totalheight=0.8\textheight]{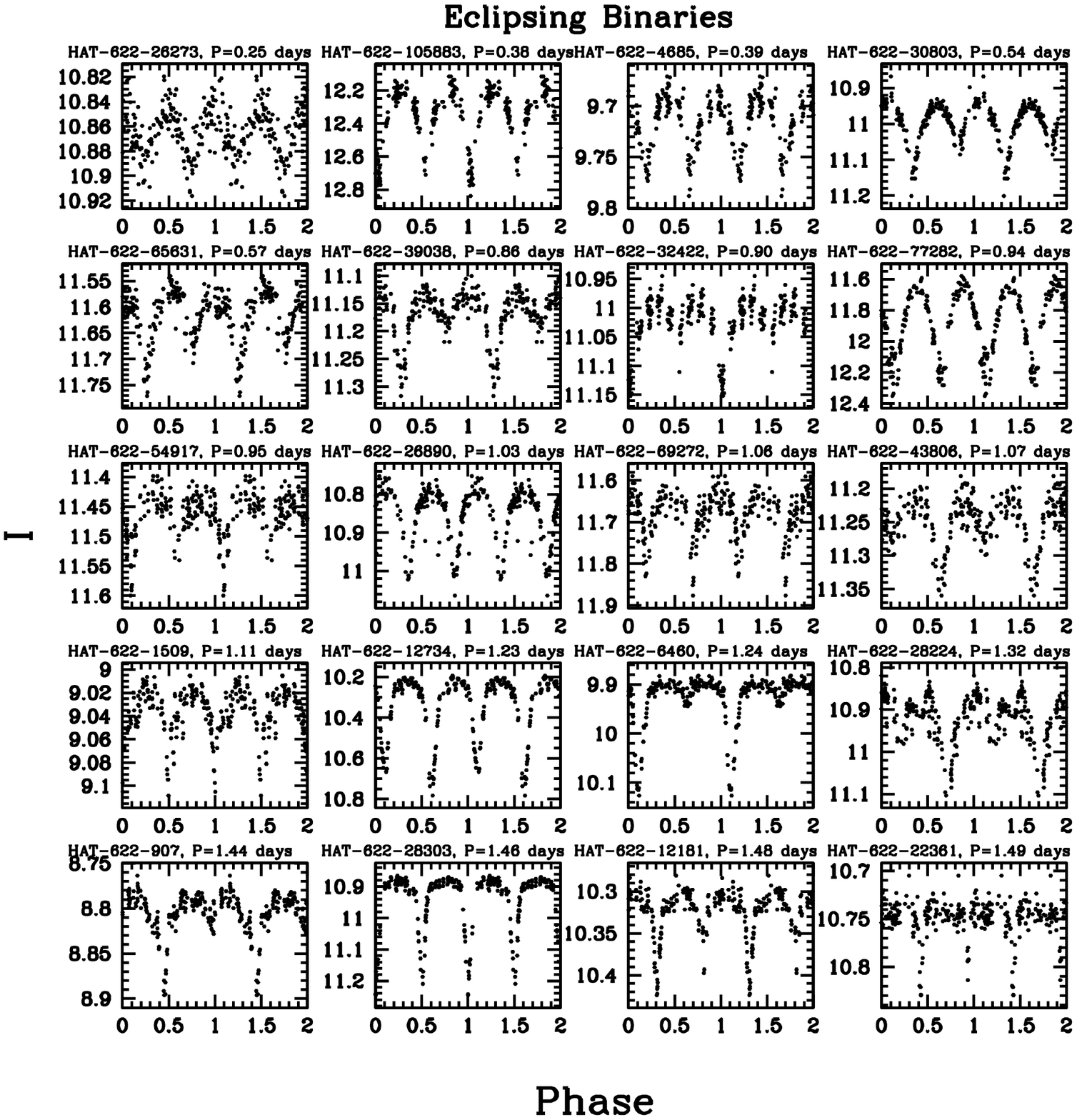}
\caption{Newly discovered eclipsing binaries, sorted by increasing period.}
\label{EM2a}
\end{figure}

\newpage
\begin{figure}[H]
\includegraphics[totalheight=0.8\textheight]{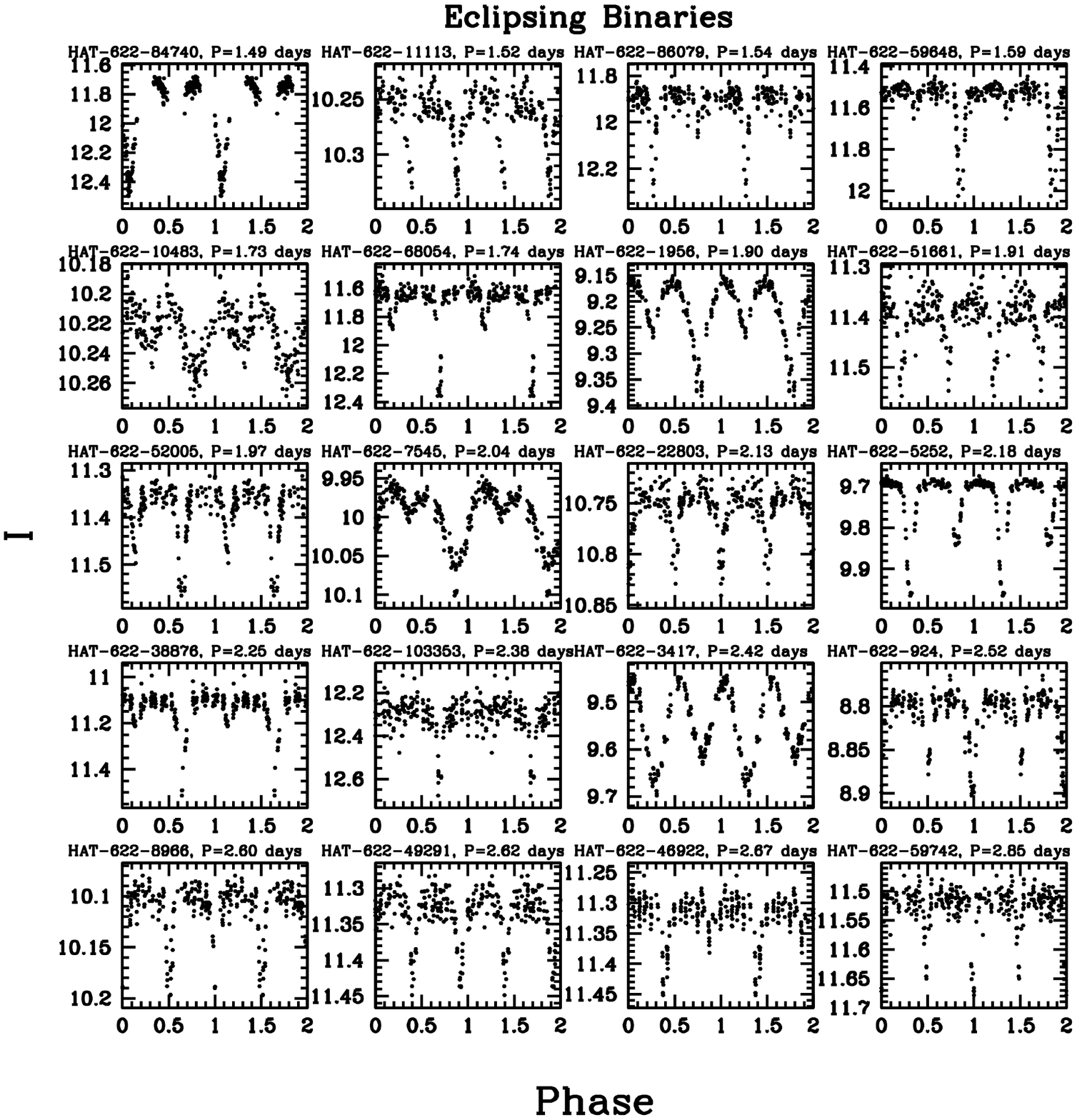}
\caption{Newly discovered eclipsing binaries (continued)}
\label{EM2b}
\end{figure}

\newpage
\begin{figure}[H]
\includegraphics[totalheight=0.8\textheight]{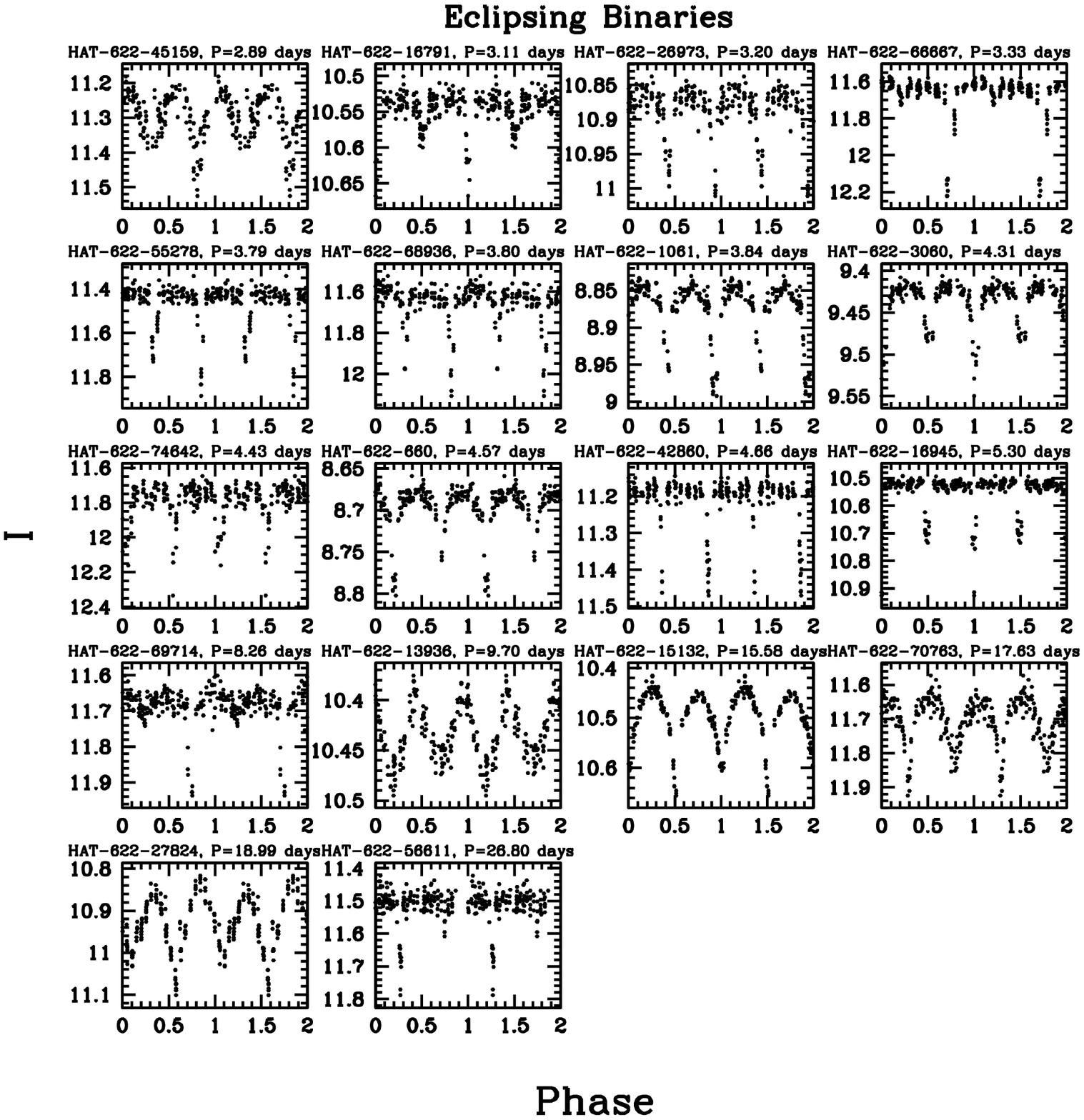}
\caption{Newly discovered eclipsing binaries (continued)}
\label{EM2c}
\end{figure}

\newpage
\begin{figure}[H]
\includegraphics[totalheight=0.8\textheight]{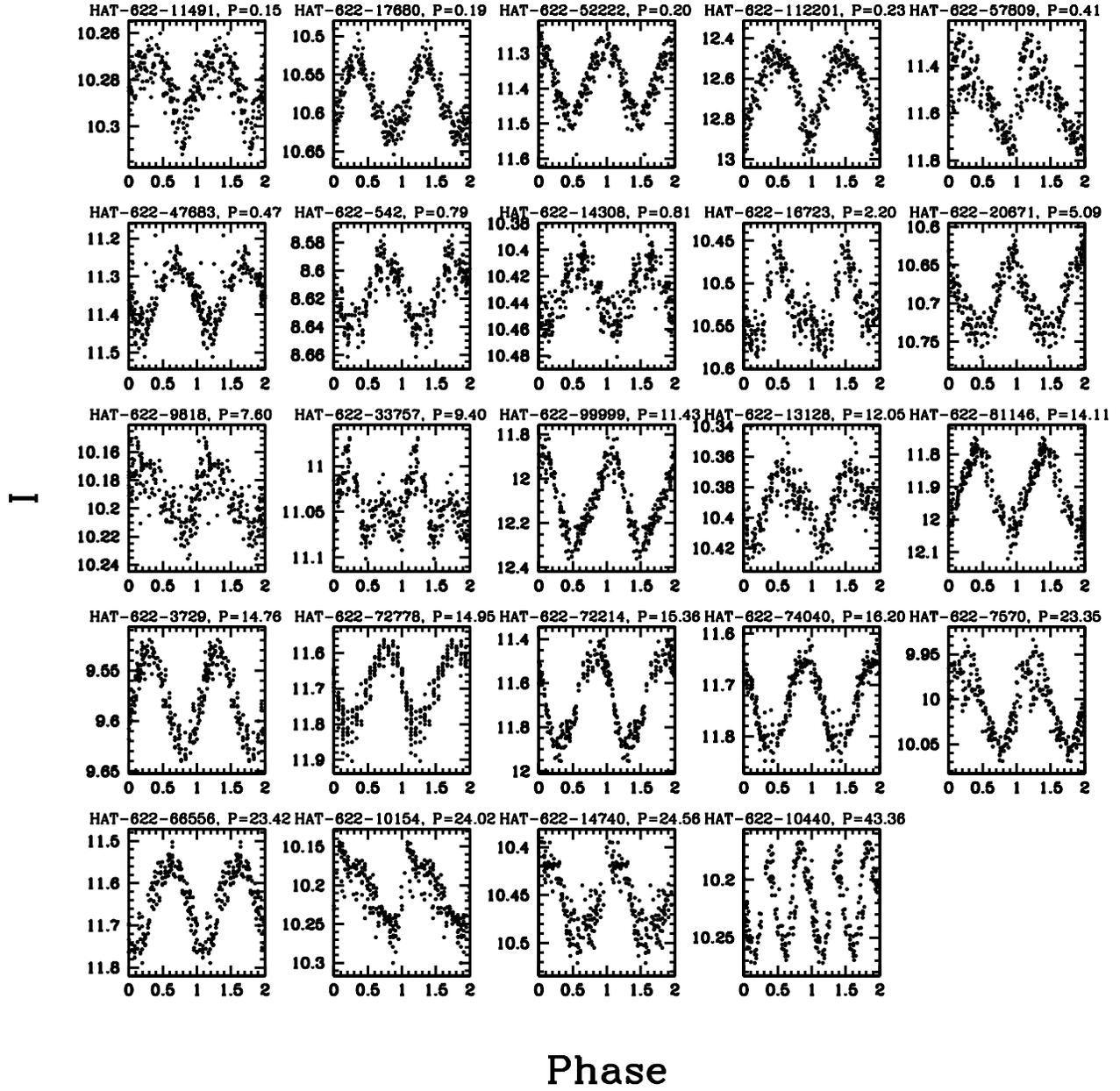}
\caption{Newly discovered periodic variable sources. Additional observations are needed to classify the type of periodicity.}
\label{PerM}
\end{figure}

\begin{appendix}
\label{appendix}

\newpage
\centering 
\begin{footnotesize}
\begin{center}
Table 1: Eclipsing Binaries
\begin{longtable}{ l l l l l l l}
\label{table:eclipsingbinaries}
\\
\hline \hline
Name & RA & DEC & I & Period & Status & 2MASS ID\\
\hline \hline 
HAT-622-159 & 18:19:36.6 & -33:01:19.1 & 8.268 & 0.847382 & Known & 18193663-3301191\\ 
HAT-622-660 & 18:06:42.3 & -32:23:03.1 & 8.69 & 4.574501 & New & 18064225-3223030\\ 
HAT-622-721 & 17:58:43.2 & -33:19:33.5 & 8.719 & 1.788577 & Known & 17584320-3319335\\ 
HAT-622-907 & 17:51:33.8 & -30:17:13.6 & 8.807 & 1.43574738 & New & 17513375-3017135\\ 
HAT-622-911 & 17:58:22.1 & -28:57:12.0 & 8.809 & 3.385964 & Known & 17582210-2857119\\ 
HAT-622-924 & 17:51:36.4 & -32:02:01.2 & 8.816 & 2.516904 & New & 17513642-3202012\\ 
HAT-622-930 & 17:59:10.4 & -27:07:43.2 & 8.818 & 0.933608 & Known & 17591044-2707432\\ 
HAT-622-1061 & 17:49:45.4 & -28:24:20.7 & 8.877 & 3.836842 & New & 17494543-2824206\\ 
HAT-622-1509 & 18:08:48.8 & -26:51:58.0 & 9.042 & 1.112803 & New & 18084884-2651579\\ 
HAT-622-1956 & 17:54:15.4 & -28:26:35.1 & 9.181 & 1.89967901 & New & 17541539-2826351\\ 
HAT-622-2503 & 17:51:51.8 & -26:58:53.1 & 9.322 & 1.881122 & Known & 17515178-2658531\\ 
HAT-622-3019 & 18:25:54.7 & -27:51:14.5 & 9.428 & 1.39468 & Known & 18255468-2751144\\ 
HAT-622-3060 & 17:52:58.9 & -28:03:40.2 & 9.437 & 4.309781 & New & 17525885-2803402\\ 
HAT-622-3417 & 17:51:43.2 & -29:31:21.5 & 9.509 & 2.42123733 & New & 17514323-2931215\\ 
HAT-622-4685 & 18:11:44.9 & -31:36:12.4 & 9.706 & 0.391400 & New & 18114486-3136123\\ 
HAT-622-5252 & 17:52:59.4 & -32:33:10.2 & 9.779 & 2.18398710 & New & 17525940-3233102\\ 
HAT-622-6460 & 18:09:52.1 & -33:31:05.6 & 9.913 & 1.24266059 & New & 18095209-3331055\\ 
HAT-622-7545 & 17:53:39.0 & -29:53:04.5 & 10.012 & 2.03666731 & New & --\\ 
HAT-622-8870 & 18:07:44.6 & -28:24:04.2 & 10.115 & 30.659727 & Known & 18074457-2824041\\ 
HAT-622-8966 & 18:14:48.1 & -32:04:04.9 & 10.123 & 2.595499 & New & --\\ 
HAT-622-9639 & 18:05:03.0 & -29:10:52.7 & 10.175 & 1.669197 & Known & 18050304-2910527\\ 
HAT-622-10483 & 17:56:01.7 & -27:50:01.3 & 10.229 & 1.72893744 & New & 17560165-2750013\\ 
HAT-622-10793 & 18:14:33.4 & -28:08:46.3 & 10.249 & 2.56192 & Known & --\\ 
HAT-622-11113 & 18:02:55.9 & -26:59:16.0 & 10.266 & 1.519982 & New & --\\ 
HAT-622-11931 & 18:18:20.0 & -28:06:30.6 & 10.314 & 6.49587 & Known & 18181995-2806305\\ 
HAT-622-12181 & 18:24:39.2 & -27:56:52.0 & 10.329 & 1.47911796 & New & --\\ 
HAT-622-12734 & 18:12:38.6 & -26:48:52.2 & 10.359 & 1.229632 & New & 18123858-2648521\\ 
HAT-622-12906 & 18:16:27.1 & -26:08:44.7 & 10.368 & 24.41 & Known & 18162712-2608446\\ 
HAT-622-13936 & 18:13:13.4 & -26:16:23.5 & 10.421 & 9.70361754 & New & 18131340-2616234\\ 
HAT-622-14496 & 18:24:57.4 & -30:24:42.6 & 10.448 & 2.2517 & Known & 18245742-3024426\\ 
HAT-622-15132 & 18:19:35.6 & -31:44:55.6 & 10.479 & 15.5760390 & New & --\\ 
HAT-622-16791 & 18:17:12.7 & -33:35:29.8 & 10.549 & 3.106143 & New & --\\ 
HAT-622-16945 & 18:05:33.6 & -26:51:43.2 & 10.555 & 5.298144 & New & 18053363-2651432\\ 
HAT-622-17137 & 17:51:09.5 & -29:21:19.8 & 10.563 & 10.664332 & Known & 17510954-2921198\\ 
HAT-622-20752 & 18:12:26.4 & -29:16:26.7 & 10.695 & 1.79772 & Known & 18122640-2916267\\ 
HAT-622-21164 & 18:11:03.9 & -30:30:56.9 & 10.709 & 2.45627004 & Known & 18110385-3030569\\ 
HAT-622-21532 & 17:54:17.5 & -29:58:48.4 & 10.721 & 2.8617 & Known & 17541752-2958483\\ 
HAT-622-22361 & 18:19:15.9 & -29:26:34.0 & 10.748 & 1.487780 & New & --\\ 
HAT-622-22803 & 18:15:06.7 & -30:30:51.2 & 10.762 & 2.130432 & New & 18150666-3030511\\ 
HAT-622-23846 & 17:59:52.5 & -28:09:33.3 & 10.793 & 5.100782 & Known & --\\ 
HAT-622-26273 & 18:17:39.4 & -27:40:11.3 & 10.863 & 0.248355 & New & 18173935-2740113\\ 
HAT-622-26342 & 18:06:51.3 & -29:00:12.0 & 10.865 & 0.3694 & Known & --\\ 
HAT-622-26890 & 18:19:56.9 & -26:00:54.8 & 10.88 & 1.031726 & New & --\\ 
HAT-622-26973 & 18:13:23.9 & -32:47:49.8 & 10.882 & 3.199622 & New & 18132385-3247498\\ 
HAT-622-27055 & 17:50:28.9 & -33:13:50.0 & 10.885 & 1.682458 & Known & 17502891-3313499\\ 
HAT-622-27318 & 17:49:10.2 & -33:24:22.1 & 10.892 & 0.79238 & Known & --\\ 
HAT-622-27824 & 18:08:16.0 & -30:10:37.4 & 10.907 & 18.994884 & New & 18081603-3010374\\ 
HAT-622-28224 & 18:08:46.4 & -33:12:37.2 & 10.917 & 1.31843018 & New & 18084639-3312371\\ 
HAT-622-28303 & 18:22:45.3 & -30:45:31.3 & 10.92 & 1.464364 & New & 18224531-3045313\\ 
HAT-622-29670 & 17:59:49.6 & -31:55:23.8 & 10.953 & 1.41833 & Known & --\\ 
HAT-622-30225 & 18:16:50.1 & -30:08:02.4 & 10.966 & 2.42366 & Known & 18165011-3008023\\ 
HAT-622-30286 & 18:23:00.3 & -32:34:03.0 & 10.968 & 2.55858224 & Known & 18230030-3234030\\ 
HAT-622-30803 & 18:20:05.2 & -29:51:25.6 & 10.98 & 0.54367619 & New & --\\ 
HAT-622-32374 & 17:52:00.4 & -27:45:08.0 & 11.016 & 1.14027 & Known & 17520035-2745080\\ 
HAT-622-32422 & 18:19:07.3 & -33:14:18.3 & 11.017 & 0.899283 & New & 18190730-3314182\\ 
HAT-622-37664 & 18:24:29.2 & -32:25:36.5 & 11.131 & 3.3517 & Known & 18242924-3225364\\ 
HAT-622-37687 & 17:50:30.4 & -29:36:48.2 & 11.132 & 1.433911 & Known & 17503040-2936482\\ 
HAT-622-38876 & 17:54:20.0 & -29:38:15.7 & 11.156 & 2.24593272 & Known & 17541996-2938157\\ 
HAT-622-39038 & 18:21:52.5 & -31:35:04.3 & 11.159 & 0.86226006 & New & --\\ 
HAT-622-40871 & 17:55:57.0 & -31:42:45.7 & 11.196 & 1.39645 & Known & --\\ 
HAT-622-41156 & 17:48:51.0 & -33:28:51.6 & 11.202 & 0.541165 & Known & --\\ 
HAT-622-42860 & 18:10:20.0 & -27:49:32.0 & 11.235 & 4.657990 & New & --\\ 
HAT-622-42871 & 17:56:55.6 & -30:55:18.2 & 11.235 & 0.395288 & Known & 17565559-3055181\\ 
HAT-622-43806 & 17:58:19.1 & -25:53:33.5 & 11.254 & 1.06862778 & New & --\\ 
HAT-622-45159 & 17:50:38.8 & -26:04:30.3 & 11.279 & 2.88845976 & New & 17503877-2604303\\ 
HAT-622-46072 & 18:02:20.8 & -30:40:17.4 & 11.295 & 0.569166 & Known & --\\ 
HAT-622-46260 & 18:16:56.2 & -28:29:29.2 & 11.298 & 0.57802 & Known & --\\ 
HAT-622-46922 & 18:16:44.0 & -31:38:56.1 & 11.311 & 2.66821923 & New & --\\ 
HAT-622-49291 & 17:59:05.1 & -29:53:10.1 & 11.353 & 2.615025 & New & --\\ 
HAT-622-51661 & 18:00:07.7 & -33:48:51.2 & 11.394 & 1.905257 & New & --\\ 
HAT-622-52005 & 18:21:50.7 & -30:09:35.4 & 11.4 & 1.974516 & New & 18215074-3009353\\ 
HAT-622-53219 & 17:58:39.2 & -33:20:38.2 & 11.42 & 0.421905 & Known & --\\ 
HAT-622-53272 & 17:54:46.7 & -30:11:24.1 & 11.421 & 30.28064 & Known & 17544667-3011240\\ 
HAT-622-54147 & 18:05:34.3 & -30:22:23.9 & 11.436 & 1.2335 & Known & --\\ 
HAT-622-54917 & 18:16:43.8 & -25:53:18.7 & 11.449 & 0.954777 & New & --\\ 
HAT-622-55278 & 18:14:08.2 & -32:22:36.6 & 11.455 & 3.786584 & New & --\\ 
HAT-622-56611 & 17:52:46.0 & -30:19:43.4 & 11.478 & 26.801783 & New & --\\ 
HAT-622-57328 & 17:52:57.5 & -28:13:31.7 & 11.489 & 1.15544669 & Known & --\\ 
HAT-622-59648 & 17:56:57.9 & -31:07:05.7 & 11.527 & 1.59332545 & Known & 17565785-3107057\\ 
HAT-622-59742 & 18:26:17.8 & -30:08:19.1 & 11.529 & 2.846187 & New & --\\ 
HAT-622-62542 & 17:50:48.9 & -28:29:23.1 & 11.575 & 1.06204 & Known & --\\ 
HAT-622-65631 & 18:15:07.1 & -28:53:20.6 & 11.625 & 0.56969497 & New & --\\ 
HAT-622-66667 & 18:23:11.0 & -28:17:25.8 & 11.642 & 3.32745807 & Known & 18231095-2817257\\ 
HAT-622-68054 & 18:07:51.2 & -33:09:38.4 & 11.664 & 1.73587480 & New & 18075117-3309384\\ 
HAT-622-68936 & 17:52:42.6 & -28:51:35.1 & 11.677 & 3.796415 & New & 17524259-2851350\\ 
HAT-622-69272 & 18:18:55.6 & -26:21:55.4 & 11.683 & 1.064236 & New & 18185558-2621553\\ 
HAT-622-69714 & 18:26:47.1 & -32:25:55.5 & 11.689 & 8.25937750 & New & --\\ 
HAT-622-70763 & 17:51:49.1 & -27:47:59.3 & 11.705 & 17.633739 & New & --\\ 
HAT-622-74308 & 18:01:59.2 & -29:13:38.2 & 11.764 & 0.73012 & Known & 18015915-2913381\\ 
HAT-622-74642 & 17:54:07.1 & -28:39:17.2 & 11.769 & 4.429529 & Known & 17540714-2839172\\ 
HAT-622-77282 & 18:08:58.9 & -27:19:03.9 & 11.813 & 0.942209 & Known & --\\ 
HAT-622-83009 & 18:16:16.6 & -33:37:48.8 & 11.907 & 28.122471 & Known & --\\ 
HAT-622-84740 & 17:50:43.2 & -28:13:14.3 & 11.937 & 1.492609 & New & --\\ 
HAT-622-86079 & 17:57:29.9 & -31:40:36.9 & 11.96 & 1.538337 & Known & 17572985-3140369\\ 
HAT-622-86466 & 17:51:33.9 & -27:52:40.2 & 11.967 & 3.209 & Known & 17513385-2752402\\ 
HAT-622-89092 & 18:11:36.2 & -30:41:08.6 & 12.013 & 5.52803544 & Known & --\\ 
HAT-622-91723 & 18:18:28.8 & -30:29:53.7 & 12.061 & 1.2989 & Known & 18182884-3029537\\ 
HAT-622-91993 & 18:20:50.2 & -27:43:02.0 & 12.066 & 1.472808 & Known & 18205019-2743019\\ 
HAT-622-103353 & 18:07:39.5 & -32:58:45.6 & 12.31 & 2.3842397 & New & --\\ 
HAT-622-104435 & 18:14:24.4 & -26:34:40.8 & 12.34 & 0.78628 & Known & --\\ 
HAT-622-105883 & 18:09:08.9 & -30:02:11.6 & 12.38 & 0.384420 & New & --\\ 
\end{longtable}
\end{center}
\end{footnotesize}

\newpage
\label{table:Periodic}
\begin{footnotesize}
\begin{center}
Table 2: Periodic variables of ambiguous classification
\begin{longtable}{ l l l l l l l}
\label{table:periodicvariables}
\\
\hline \hline
 Name & RA & DEC & I & Period & Status & 2MASS ID\\
\hline \hline 
HAT-622-542 & 17:51:06.8 & -32:18:27.7 & 8.616 & 0.79162542 & New & 17510684-3218276\\ 
HAT-622-3201 & 17:54:17.6 & -26:45:29.5 & 9.465 & 5.7485 & Known & 17541760-2645294\\ 
HAT-622-3729 & 17:50:24.8 & -30:26:24.3 & 9.564 & 14.75985462 & New & 17502475-3026242\\ 
HAT-622-7570 & 17:53:48.1 & -28:04:14.6 & 10.014 & 23.353579 & New & 17534805-2804145\\ 
HAT-622-9001 & 18:20:18.4 & -33:20:08.4 & 10.126 & 5.8241 & Known & 18201843-3320084\\ 
HAT-622-9818 & 18:04:48.9 & -29:34:18.4 & 10.186 & 7.59996876 & New & 18044891-2934183\\ 
HAT-622-10154 & 18:01:11.7 & -29:32:02.5 & 10.208 & 24.021954 & New & --\\ 
HAT-622-10359 & 17:56:21.5 & -27:37:55.8 & 10.221 & 11.60971 & Known & 17562148-2737557\\ 
HAT-622-10440 & 18:24:23.1 & -30:33:17.6 & 10.226 & 43.357149 & New & 18242310-3033176\\ 
HAT-622-11491 & 17:58:45.8 & -28:20:11.4 & 10.288 & 0.15498803 & New & --\\ 
HAT-622-12341 & 17:58:54.9 & -28:58:35.9 & 10.338 & 25.687367 & Known & --\\ 
HAT-622-13128 & 18:14:22.4 & -32:46:12.1 & 10.38 & 12.04894357 & New & --\\ 
HAT-622-14308 & 17:52:17.4 & -27:39:42.2 & 10.439 & 0.80660440 & New & 17521741-2739422\\ 
HAT-622-14740 & 18:07:02.9 & -33:03:12.9 & 10.46 & 24.55724684 & New & 18070294-3303128\\ 
HAT-622-15750 & 18:17:19.3 & -33:21:03.5 & 10.505 & 14.86922 & Known & --\\ 
HAT-622-16723 & 17:51:36.1 & -30:14:01.7 & 10.546 & 2.19533409 & New & --\\ 
HAT-622-17680 & 17:59:01.3 & -26:43:00.3 & 10.584 & 0.19258022 & New & 17590134-2643002\\ 
HAT-622-20671 & 18:06:27.6 & -26:33:09.7 & 10.692 & 5.091530 & New & --\\ 
HAT-622-22364 & 17:52:28.4 & -26:42:12.5 & 10.748 & 0.670304 & Known & 17522840-2642124\\ 
HAT-622-24311 & 18:03:57.9 & -29:56:59.8 & 10.807 & 0.0844 & Known & 18035790-2956597\\ 
HAT-622-28020 & 18:20:46.6 & -29:48:49.0 & 10.912 & 15.04708965 & Known & --\\ 
HAT-622-28201 & 17:54:14.6 & -27:28:18.9 & 10.917 & 16.67188 & Known & 17541464-2728189\\ 
HAT-622-32968 & 18:15:24.6 & -31:04:10.5 & 11.03 & 16.20146 & Known & --\\ 
HAT-622-33757 & 18:11:27.7 & -30:02:11.9 & 11.048 & 9.40194621 & New & 18112766-3002119\\ 
HAT-622-34676 & 18:03:59.0 & -30:11:11.8 & 11.07 & 0.49126 & Known & --\\ 
HAT-622-37905 & 18:04:08.0 & -33:46:54.0 & 11.137 & 11.59407 & Known & 18040802-3346540\\ 
HAT-622-37992 & 18:04:41.1 & -28:56:04.9 & 11.138 & 0.192918 & Known & 18044112-2856048\\ 
HAT-622-40536 & 18:16:50.3 & -27:47:52.7 & 11.189 & 0.44087 & Known & 18165028-2747526\\ 
HAT-622-47683 & 18:24:00.3 & -25:58:54.5 & 11.324 & 0.46782932 & New & 18240025-2558544\\ 
HAT-622-52222 & 18:01:36.5 & -32:28:42.3 & 11.403 & 0.20325365 & New & 18013650-3228423\\ 
HAT-622-57809 & 18:11:13.6 & -29:42:40.3 & 11.497 & 1.23919507 & New & --\\ 
HAT-622-60112 & 18:19:46.4 & -27:09:28.2 & 11.535 & 13.91738 & Known & --\\ 
HAT-622-65557 & 18:14:08.5 & -27:49:31.4 & 11.624 & 21.78503371 & Known & --\\ 
HAT-622-66556 & 17:56:01.5 & -33:10:46.5 & 11.64 & 23.42059316 & New & --\\ 
HAT-622-72214 & 18:09:32.7 & -27:42:15.5 & 11.729 & 15.36038098 & New & --\\ 
HAT-622-72612 & 18:10:24.0 & -32:28:10.5 & 11.736 & 22.47432836 & Known & --\\ 
HAT-622-72778 & 18:02:37.2 & -28:34:38.3 & 11.739 & 14.95494757 & New & 18023717-2834382\\ 
HAT-622-74040 & 17:58:22.2 & -29:58:52.8 & 11.759 & 16.19772235 & New & 17582218-2958528\\ 
HAT-622-81146 & 18:25:53.0 & -33:27:43.3 & 11.877 & 14.11048761 & New & 18255302-3327433\\ 
HAT-622-82362 & 18:20:26.6 & -27:45:41.7 & 11.897 & 0.52061004 & Known & --\\ 
HAT-622-84951 & 17:58:44.2 & -33:11:54.3 & 11.941 & 18.319 & Known & --\\ 
HAT-622-94433 & 18:23:19.1 & -32:25:34.4 & 12.113 & 0.58133582 & Known & --\\ 
HAT-622-96670 & 18:18:09.0 & -32:59:17.4 & 12.158 & 12.13572364 & Known & 18180901-3259173\\ 
HAT-622-99999 & 17:57:36.1 & -26:51:13.8 & 12.23 & 11.43233439 & New & 17573613-2651138\\ 
HAT-622-104492 & 18:22:56.3 & -28:04:38.7 & 12.342 & 14.97269335 & Known & --\\ 
HAT-622-112201 & 18:22:22.0 & -33:19:16.2 & 12.64 & 0.22875900 & New & --\\ 
\end{longtable}
\end{center}
\end{footnotesize}

\label{table:knownPeriodic}
\begin{footnotesize}
\begin{center}
Table 3: Properties of known periodic variables with detailed matches in GCVS or ASAS. Some changes to the GCVS classifications made for brevity: Semi-regular pulsating stars have been listed to SRPS, Classical Cepheids (delta Cep) as ($\delta$ Cep), and Cepheid variable stars as Cepheid 
\begin{longtable}{ l l l l l l l}
\label{table:knownperiodicvariables}
\\
\hline \hline
HAT ID & ASAS ID & ASAS Class. & ASAS P. & GCVS & GCVS Class. & CCVS P. \\
\hline \hline
HAT-622-159 & 181936-3301.4 & EC/ESD & 0.847382 & - & - & - \\
HAT-622-721 & 175843-3319.5 & ED & 1.788577 & V1721 Sgr & EA & 1.7886 \\
HAT-622-911 & 175822-2857.2 & ED & 3.385964 & - & - & - \\
HAT-622-930 & 175910-2707.7 & EC/ESD & 0.933608 & - & - & - \\
HAT-622-2503 & 175152-2658.9 & ED & 1.881122 & - & - & - \\
HAT-622-3019 & 182555-2751.2 & ED & 1.39468 & - & - & - \\
HAT-622-3201 & 175417-2645.4 & DECP-FU & 5.7485 & V773 Sgr & ($\delta$ Cep) & 5.7501 \\ 
HAT-622-8870 & 180745-2824.1 & ED & 30.659727 & - & - & - \\
HAT-622-9001 & 182019-3320.1 & DCEP-FU & 5.8241 & - & - & - \\
HAT-622-9639 & 180503-2910.9 & ED/ESD & 1.669197 & - & - & - \\
HAT-622-10359 & 175621-2738.0 & DCEP-FU & 11.60971 & - & - & - \\
HAT-622-10793 & 181434-2808.8 & ESD/ED & 2.56192 & - & - & - \\
HAT-622-11931 & 181820-2806.5 & ED: & 6.49587 & - & - & - \\
HAT-622-12341 & - & - & - & V5218 Sgr & SRPS & 25.00 \\
HAT-622-12906 & 181627-2608.8 & MISC & 12.205 & - & - & - \\
HAT-622-14496 & 182457-3024.7 & ESD/ED & 2.2517 & V5535 Sgr & EB & 2.2517 \\
HAT-622-15750 & 181719-3321.1 & DCEP-FU & 14.86922 & - & - & - \\
HAT-622-17137 & 175110-2921.3 & ED & 10.664332 & - & - & - \\
HAT-622-20752 & 181226-2916.4 & ED/ESD & 1.79772 & - & - & - \\
HAT-622-21164 & - & - & - & V1178 Sgr & EA & 2.458807 \\
HAT-622-21532 & 175418-2958.9 & ESD & 2.8617 & - & - & - \\
HAT-622-22364 & 175228-2642.2 & DCEP-FO & 0.670304 & V767 Sgr & RR Lyr & 0.6702 \\
HAT-622-23846 & - & - & - & V789 Sgr & EA & 2.55234 \\
HAT-622-24311 & 180358-2957.1 & DSCT & 0.0844 & V5505 Sgr & delta Sct & 0.08493 \\
HAT-622-26342 & 180650-2900.2 & EC/DSCT/ESD & 0.3694 & - & - & - \\
HAT-622-27055 & 175029-3313.8 & ED & 1.682458 & - & - & - \\
HAT-622-27318 & 174910-3324.3 & ED/ESD & 0.79238 & - & - & - \\
HAT-622-28020 & - & - & - & V741 Sgr & W Vir & 15.156 \\
HAT-622-28201 & 175414-2728.2 & MISC & 16.67188 & - & - & - \\
HAT-622-29670 & 175950-3155.3 & ED/ESD & 1.41833 & V994 Sgr & EA & 1.4183 \\
HAT-622-30225 & 181650-3008.0 & EC & 2.42366 & - & - & - \\
HAT-622-30286 & - & - & - & V2537 Sgr & EA & 2.5581 \\
HAT-622-32374 & 175200-2745.1 & ESD/ED & 1.14027 & - & - & - \\
HAT-622-32968 & 181525-3104.2 & DCEP-FU & 16.20146 & - & - & - \\
HAT-622-34676 & 180359-3011.1 & EC/ESD & 0.49126 & - & - & - \\
HAT-622-37664 & 182429-3225.6 & ESD & 3.3517 & - & - & - \\
HAT-622-37687 & - & - & - & V761 Sgr & EA & 1.4334 \\
HAT-622-37905 & 180408-3347.0 & DECP-FU & 11.59407 & V1008 Sgr & Variable Star & 11.5918 \\
HAT-622-37992 & 180441-2856.2 & DSCTr/DSCT & 0.192918 & - & - & - \\
HAT-622-40536 & 181650-2747.9 & RRAB & 0.44087 & V1182 Sgr & RR Lyr & 0.4409 \\
HAT-622-40871 & 175557-3142.8 & ED & 1.39645 & V714 Sco & EA & 0.6982 \\
HAT-622-41156 & 174851-3328.9 & ESD/EC & 0.541165 & - & - & - \\
HAT-622-42871 & 175656-3055.3 & EC & 0.395288 & - & - & - \\
HAT-622-46072 & 180221-3040.4 & EC & 0.569166 & - & - & - \\
HAT-622-46260 & 181656-2829.4 & EC & 0.57802 & - & - & - \\
HAT-622-53219 & 175839-3320.6 & EC - & 0.421906 & - & - & - \\
HAT-622-53272 & 175447-3011.5 & MISC & 15.14032 & V712 Sco & EA & 30.3050 \\
HAT-622-54147 & 180534-3022.5 & ED & 1.2335 & - & - & - \\
HAT-622-57328 & - & - & - & V1603 Sgr & EA & 3.3327 \\
HAT-622-60112 & 181947-2709.4 & DCEP-FU & 13.91738 & V1185 Sgr & Cepheid & 13.9125 \\
HAT-622-62542 & 175049-2829.6 & ESD & 1.06204 & - & - & - \\
HAT-622-65557 & - & - & - & V1181 Sgr & W Vir & 21.315 \\
HAT-622-66667 & - & - & - & V1603 Sgr & EA & 3.3327 \\
HAT-622-72612 & - & - & - & V2505 Sgr & W Vir & 22.5569 \\
HAT-622-74308 & 180159-2913.6 & EC & 0.73012 & - & - & - \\
HAT-622-74642 & - & - & - & V5559 Sgr & EA & 4.4361 \\
HAT-622-77282 & 180859-2719.1 & EC & 0.942342 & - & - & - \\
HAT-622-82362 & - & - & - & VV1288 Sgr & RR Lyr & 0.5205 \\
HAT-622-83009 & - & - & - & V1834 Sgr & W Vir & 14.0036 \\
HAT-622-84951 & 175843-3311.9 & DCEP-FU/EC & 18.319 & - & - & - \\
HAT-622-86466 & 175134-2752.6 & ED & 3.209 & - & - & - \\
HAT-622-89092 & - & - & - & V2507 Sgr & E & 5.5095 \\
HAT-622-91723 & - & - & - & V1184 Sgr & EA &  1.2989 \\
HAT-622-91993 & - & - & - & V2529 Sgr & EB & 0.7363 \\
HAT-622-94433 & - & - & - & V1188 Sgr & RR Lyr & 0.5813 \\
HAT-622-96670 & - & - & - & V2944 Sgr & Variable Star & 12.101 \\
HAT-622-104435 & 181425-2634.7 & EC & 0.78628 & - & - & - \\
HAT-622-104492 & - & - & - & V1187 Sgr & W Vir & 15.115 \\
\end{longtable}
\end{center}
\end{footnotesize}

\end{appendix}

\end{document}